\documentclass[aps,prd,showpacs,twocolumn]{revtex4}
\usepackage{amsmath}
\usepackage{amssymb}

\DeclareFontFamily{OT1}{rsfs}{}
\DeclareFontShape{OT1}{rsfs}{m}{n}{<-7> rsfs5 <7-10> rsfs7 <10->rsfs10}{} 
\DeclareMathAlphabet{\mycal}{OT1}{rsfs}{m}{n}

\newcommand{\g}{\mycal G}

\newcommand{\ag}{\mathfrak{a}}
\newcommand{\bg}{\mathfrak{b}}
\newcommand{\cg}{\mathfrak{c}}
\newcommand{\dg}{\mathfrak{d}}
\newcommand{\tr}{{\rm\bf tr}}

\begin{document}

\title{The Einstein-Yang-Mills equations from Bianchi identities}

\author{Christian G.~B\"ohmer}
\email{c.boehmer@ucl.ac.uk} 
\affiliation{Department of Mathematics, University College London,
             Gower Street, London, WC1E 6BT, UK}
\affiliation{Institute of Cosmology \& Gravitation,
             University of Portsmouth, Portsmouth PO1 2EG, UK}

\author{Luca Fabbri}
\email{Luca.Fabbri@bo.infn.it} 
\affiliation{Theory Group, INFN -- Department of Physics, 
             University of Bologna, Via Irnerio 46,
             C.A.P. 40126, Bologna, Italy}

\date{\today}

\begin{abstract}
We suggest a novel extension to the Kaluza-Klein scheme that allows us to obtain 
consistently all $SU(n)$ Einstein-Yang-Mills theories. This construction is based
on allowing the five-dimensional spacetime to carry some non-vanishing torsion;
however, the four-dimensional spacetime remains intrinsically torsion-free. 
\end{abstract}

\pacs{04.50.+h, 11.25.Mj}
\maketitle

\section{Introduction}
The great success of Einstein gravity opens deeper philosophical issues: 
If we assume that any physical theory should be generally covariant, then 
it is difficult to understand why all the other interactions are not described 
side by side with gravity in the framework of General Relativity. 

Kaluza~\cite{Kaluza:1921tu} and Klein~\cite{Klein:1926tv} proposed to pursue 
the electro-gravitational unification by enlarging the spacetime to a $5$-dimensional 
spacetime, in which the extra dimension was thought to carry the information concerning 
electromagnetism (for general reviews, see e.g.~\cite{Lee:1985qp,Appelquist:1987}). 
By compactification of this fifth dimension, the resulting 
theory could describe properly the Einstein-Maxwell theory;
for some critical account see~\cite{Fabbri:2002ud}. Such 
compactifications became of great interest in the context of higher 
dimensional theories. These higher dimensional spacetimes are 
important due to the fact that they can host supergravity or 
(super) string theories.

On the other hand, Weyl~\cite{Weyl:1918ib} in 1918 tried to extend General Relativity 
by considering a conformally rescaled metric $g_{ab} \rightarrow \phi g_{ab}$. 
The covariant derivative of the rescaled metric does not necessarily vanish and 
leads to non-metricity which was considered to be related the electromagnetic 
potential. Weyl, as many others later, attempted to include electromagnetism 
in a theory already containing gravity, hoping to accomplish their formal 
unification. Details on the history of such unified field theories can be 
found in~\cite{Goenner:2004se}. Although Weyl's original idea had many problems 
of its own, it later went on to become the key ingredient for gauge field 
theories, the building blocks of the Standard Model of particle physics. 

Once tensor calculus was recognized as the geometrical structure for 
General Relativity, the symmetry of the connection was established in 
order to represent the Principle of Equivalence in a 4-dimensional 
spacetime. On this background, the simplest differential identities of 
the theory, namely the Bianchi identities, implied the existence of 
conservation laws. This match between geometrical (Einstein tensor) and 
physical (energy-momentum tensor) quantities was the key point to satisfy
Mach's principle on the origin of masses, which motivated the form of
the gravitational field equations.

Ever since Weyl's approach, similar unification ideas were proposed. During the 
1920s, Cartan, see e.g.~\cite{Cartan:1923} and later works, tried to find the room 
necessary to incorporate electromagnetism into General Relativity by removing 
the assumption on the symmetry of the connection, thereby introducing the concept of 
torsion~\cite{Hehl:1976kj,Sabbata:1994}. In these torsion theories the 
connection contains a skew-symmetric part, the Cartan or torsion tensor, 
which plays the role of a physical quantity. Often the torsion of the
spacetime is related to the spin of the particle, so that mass couples
to the curvature of spacetime whereas spin couples to its torsion.

There are also some recent attempts to understand the neutrino in the
context of geometries with torsion. The Weyl Lagrangian turns out to
be equivalent to a purely geometrical Lagrangian~\cite{Vassiliev:2006ux,Vassiliev:2007}.

Yet another way is possible. One can take into account both the possibility 
of a higher dimensional spacetime and the possibility that the corresponding 
higher dimensional torsion is non-zero~\cite{Kalinowski:1981,German:1985tj,Kohler:2000,Oh:1989}. 
The geometrical approach used by Einstein can easily be extended in order 
to consider higher dimensional manifolds and the corresponding
Bianchi identities. These will be used to define higher dimensional 
conservation laws.

For a thorough analysis of the Kaluza-Klein scheme in the presence of 
torsion and gauge fields, we refer the reader to Kalinowski~\cite{Kalinowski:1981}
and references therein. In general, for higher order gauge fields, a
$(4+\dim(\g))$ dimensional manifold has been considered in the past.

Another interesting approach in that direction is the 5-dimensional projective
unified theory of Schmutzer~\cite{Schmutzer,Schmutzer-deSabbata} where the effective 
4-dimensional theory is obtained by using a postulated projection tensor
to reduce the higher-dimensional theory. However, it has been shown that 
this theory is in disagreement with experiments unless modified by assuming 
a totally antisymmetric torsion tensor~\cite{deSabbata-Gasperini}.
Note that we will consider the 5-dimensional Bianchi identities in order 
to derive the effective theory.

So far, however, all approaches to extend the original Kaluza-Klein scheme 
in order to also describe higher order gauge fields with semi-simple and 
compact Lie group $\g=SU(n)$, have not yielded promising results. The main reason 
for this is that the gauge potentials and field strengths have an internal 
group index. This makes it difficult to add a geometrical quantity linear 
in the potential, as in the $U(1)$ case. 

In the present work, we will consider a framework in which the higher 
dimensional torsion is non-vanishing but the 4-dimensional spacetime
is torsion-free. Since we are not considering fermionic matter sources 
such an assumption is well motivated. Moreover we will introduce a set of 
auxiliary vector fields $u^{\ag}$, which span a real $\dim(\g)$-dimensional 
vector space $\mathbb{R}^{\dim(\g)}$. We show that this approach will 
enable us to obtain the complete $SU(n)$ Einstein-Yang-Mills equations.

\section{Kaluza-Klein reduction with torsion}

Let us recall the relevant geometrical equations necessary to compactify
a 5-dimensional spacetime to a 4-dimensional one. Assume that the 
5d manifold is allowed to be torsioned, however, the 4d manifold is assumed
to be torsion-free, in agreement with all observations so far.
Let the metric be of the following form
\begin{align}
      ds^2 = \eta_{mn} e^m \otimes e^n = 
      \eta_{ab} e^a \otimes e^b - e^5 \otimes e^5, \\
      e^5 = dx^z, \qquad de^5 = 0,
\end{align}
where the indices of the middle of the alphabet $(m,n,\ldots)$ are complete 
spacetime indices and letters from the beginning of the alphabet $(a,b,\ldots)$ 
take values $1,2,3,4$ and letters from the end $(z)$ take the value $5$. We 
choose $\eta_{mn} = {\rm diag}(+1,-1,-1,-1,-1)$, 
see~\cite{Grumiller:2002nm,Balasin:2005}.

The following relations define the torsion 2-form to be
\begin{align}
      \tilde{T}^m = D e^m = de^m + \omega^m{}_n \wedge e^n.
      \label{eq:t}
\end{align} 
In order to further simplify the torsion contributions due to the presence of
the 5d torsion, we moreover assume $\omega^z{}_{az} = 0$ (the $z$-component of 
the connection 1-form $\omega^z{}_{a} = 0$). According to our metric, we may 
split the index $m=(a,z)$ which yields
\begin{align}
      \tilde{T}^a &= De^a = de^a + \omega^a{}_b \wedge e^b + \omega^a{}_z \wedge e^z,
      \label{eq:ta} \\
      \tilde{T}^z &= De^z = \omega^z{}_a \wedge e^a.
      \label{eq:tz}
\end{align}
The first torsion 2-form $\tilde{T}^a$ can further be split into an intrinsically
4d torsion part $T^a$ and another part coming from the additional 
spacetime structure of the fifth dimension. Since we assume the 4d spacetime to be 
torsion free, we take the intrinsic torsion part to vanish, $T^a=0$.
This simply states that the 4d spin-connection can be computed from the 
vielbein $e^a$, $\omega^a{}_b=\omega^a{}_b (e)$, whereas the spin-connection 
components $\omega^a{}_z$ yield an additional field contribution to the 4d 
spacetime
\begin{align}
      \tilde{T}^a &= \omega^a{}_z \wedge e^z, 
      \label{eq:ta2} \\
      \tilde{T}^z &= \omega^z{}_b \wedge e^b.
      \label{eq:tz2}
\end{align}
We expand $\omega^z{}_b$ in terms of basis vielbeins and write Eq.~(\ref{eq:tz2}) 
to arrive at
\begin{align}
      \tilde{T}^z = \omega^z{}_{ba}\, e^a \wedge e^b = F_{ab}\, e^a \wedge e^b,
      \label{eq:tz3}
\end{align}
where we introduced the 2-form $F$. The components of $F$ are those of the 
spin-connection component of the additional dimension $F_{ab}=\omega^z{}_{ba}$. 
Therefore, equation~(\ref{eq:tz3}) is simply
\begin{align}
      \tilde{T}^z = F_{ab}\, e^a \wedge e^b = F.
      \label{eq:tz4}
\end{align}
From the definition of $F_{ab}$ we can now easily find the analog
expression involving $\omega^b{}_z$
\begin{align}
      F_{ab} = \omega^z{}_{ba} = -\omega_b{}^z{}_a = \omega_{bza},
\end{align}
which after raising the index $b$ simply yields
\begin{align}
      \omega^b{}_{za} = F_a{}^b, \qquad \omega^b{}_z = F_a{}^b e^a.
\end{align}
Hence, from~(\ref{eq:ta2}) the torsion 2-form $\tilde{T}^a$ becomes
\begin{align}
     \tilde{T}^a = F_b{}^a e^b \wedge e^z.
\end{align}

The curvature 2-form is defined by
\begin{align}
      \tilde{R}^m{}_n = (D^2)^m{}_n = d\omega^m{}_n + \omega^m{}_l \wedge \omega^l{}_n
\end{align}
which according to the metric splits as follows
\begin{align}
      \tilde{R}^a{}_b &= d\omega^a{}_b + \omega^a{}_c \wedge \omega^c{}_b 
      + \omega^a{}_z \wedge \omega^z{}_b, \\
      \tilde{R}^a{}_z &= d\omega^a{}_z + \omega^a{}_b \wedge \omega^b{}_z, \\
      \tilde{R}^z{}_a &= d\omega^z{}_a + \omega^z{}_b \wedge \omega^b{}_a,
\end{align}
The last term of $\tilde{R}^a{}_b$ can be re-written in terms of $F$ and becomes
\begin{align}
      \omega^a{}_z \wedge \omega^z{}_b = F_c{}^a F_{db}\, e^c \wedge e^d
\end{align}
Since the first two terms of $\tilde{R}^a{}_b$ are the intrinsic 4d curvature
of that spacetime we find
\begin{align}
      \tilde{R}^a{}_b = R^a{}_b +  F^a{}_c F_{bd}\, e^c \wedge e^d
      \label{eq:c}
\end{align}
For the other curvature 2-form components we get
\begin{align}
      \tilde{R}^a{}_z &= D F_c{}^a \wedge e^c,
      \label{eq:r1}\\
      \tilde{R}^z{}_a &= D F_{ca} \wedge e^c,
      \label{eq:r2}
\end{align}
where we took into account $\tilde{R}^a{}_z=-\tilde{R}_z{}^a=\tilde{R}^{za}$. 
From these one can compute the Ricci 1-forms and the Ricci scalar and the resulting 
Einstein 1-form which yields the following (anholonomic) components of the Einstein 
tensor
\begin{align}
      \tilde{G}_{ab} &= G_{ab} - F_{ac} F^c{}_b - \frac{1}{2} F_{cd} F^{cd} \eta_{ab},\\
      \tilde{G}_{zb} &= \nabla_a F_b{}^a,\\ 
      \tilde{G}_{zz} &= \frac{1}{2} R + \frac{1}{2} F_{cd} F^{cd}.
\end{align}

\section{Bianchi identities and field equations}
\label{bianchi}

The contracted third Bianchi identity ($D\tilde{R}^a{}_b=0$) yields the 
vanishing of the covariant derivative 
of the Einstein tensor, which is one of the most important aspects of the 
gravitational field equations. The conservation of energy-momentum can be 
regarded solely as a consequence of the geometry of spacetime. 

In our $4+1$ approach, the contracted Bianchi identity splits into two pieces
\begin{align}
      \nabla^b G_{ab} &= F_{ac} (\nabla^b F^c{}_b),
      \label{eq:bi1}\\
      \nabla^b (\nabla_a F_b{}^a) &= 0.
      \label{eq:bi2}
\end{align}
These equations can be re-written in a more familiar form. Since we assume that 
the 4d spacetime is intrinsically torsion-free and moreover that $F_{az}=0$, 
the following (purely) geometrical relation is valid
\begin{align}
      \nabla_a F_{bc} + \nabla_b F_{ca} + \nabla_c F_{ab} = 0,
\end{align}
which implies that
\begin{align}
      \frac{1}{4} \nabla_c (F_{ab} F^{ab}) = -F^{ab} \nabla_a F_{bc}.
      \label{eq:bi3}
\end{align}
If we now use equations~(\ref{eq:bi2}) and~(\ref{eq:bi3}) to rewrite the
right-hand side of~(\ref{eq:bi1}), the following is found
\begin{align}
      \nabla^b G_{ab} = \nabla^b \Bigl(F_{ac} F^c{}_b 
      + \frac{1}{4}\eta_{ab} F_{cd} F^{cd} \Bigr).
      \label{eq:bi4}
\end{align}
This immediately implies the identification of the right-hand side with the
energy-momentum tensor of the electromagnetic field
\begin{align}
      T_{ab} = F_{ac} F^c{}_b + \frac{1}{4}\eta_{ab} F_{cd} F^{cd},\qquad
      U(1)\ {\rm gauge\ theory}.
      \label{eq:bi5}
\end{align}

Before proceeding further and generalizing the above the arbitrary $SU(n)$
gauge groups, it is important to show the consistency of our above result
with the recent criticism put forward by Hehl~\cite{Hehl:2006rv,Hehl2,Hehl3}. Hehl
argued that the electromagnetic field strength cannot be simply related
to the torsion of spacetime in a 4d spacetime. The main
geometrical argument behind that is based on the fact that the torsion
1-form decomposes into three irreducible pieces and that one must relate
the field strength to any of these pieces rather than some suitable
combination. In general, the components of the torsion tensor decompose 
into irreducible components (a vector part, an axial vector part and a 
tensor part) as
\begin{align}
      T^m &= {}^{(1)}T^m + {}^{(2)}T^m + {}^{(3)}T^m, \\
      \frac{1}{2} n^2 (n-1) &= \frac{1}{3}n(n^2-4) + n + \frac{1}{6}n(n-1)(n-2),
\end{align}
which in four dimensions becomes
\begin{align}
      24 = 16 + 4 + 4.
      \label{eq:dec4}
\end{align}
Since no irreducible piece has 6 independent components, none can directly
account for the Abelian field strength. In our approach, however, we are
considering a 5d space, for which the decomposition reads
\begin{align}
      50 = 35 + 5 + 10.
      \label{eq:dec5}
\end{align}
If we furthermore take into account that we assumed the intrinsically 4d spacetime to be 
torsion-free, then the resulting 5d torsion tensor has less allowed components,
namely the difference of~(\ref{eq:dec5}) and (\ref{eq:dec4}), and we find
the following number of allowed irreducible components
\begin{align}
      (50-24) &= (35-16) + (5-4) + (10-4),\nonumber \\ 
      26 &= 19 + 1 + 6.
      \label{eq:deckk}
\end{align}
Therefore, in the torsioned Kaluza-Klein scheme, there exists one 
(and only one) irreducible component of the higher dimensional torsion
tensor that has 6 independent components, the axial torsion vector part,
if the intrinsic 4d torsion is assumed to be zero. It is worth noting
that it has been argued recently~\cite{Fabbri:2006xq}, that the torsion 
tensor should be totally antisymmetric, quite independent of the model, 
by assuming a strong metricity argument. This observation, as well as 
our result, are also in agreement with the suggested 
modifications~\cite{deSabbata-Gasperini} in the context of projective 
unified theory~\cite{Schmutzer,Schmutzer-deSabbata}.

So far $F_{ab}$ may only account for the Abelian $U(1)$ field
strength. However, it is our aim to find the Einstein-Yang-Mills equations
for any compact, semi-simple gauge group $\g$. The Lie algebra of 
$\g$ is characterized by the commutation relations 
$[T_{\ag},T_{\bg}]=i f_{\ag\bg}{}^{\cg} T_{\cg}$, $\ag,\bg,\cg =1,\ldots,\dim(\g)$,
where $T_{\ag}$ are the generators of the Lie algebra.
Furthermore we have the relations $\tr(T_{\ag}) = 0,$ for all $\ag$ and
$\tr(T_{\ag} T_{\bg})=K\delta_{\ag\bg}$, where $K>0$ is a normalization factor.

We have up to now assumed that the geometrical quantities are not Lie algebra
valued and therefore we cannot simply replace $F_{ab}$ by $F_{ab}^{\ag}T_{\ag}$.
Neither can we use the trace $\tr(F_{ab}^{\ag}T_{\ag})$,
since $\tr(T_{\ag})=0$ and therefore this term would also vanish.
On the other hand, due to the contractions to get the Einstein tensor, 
the torsion and hence the terms with $F_{ab}$ would appear quadratically,
as expected from the known form of the Yang-Mills energy-momentum tensor.
Hence, we cannot use a term quadratic in the field strength, as this would 
yield a quartic term. However, these problems can be circumvented by 
introducing an auxiliary field $U^{\ag}$ in the following way
\begin{align}
      F_{ab} \rightarrow \tr \left( F_{ab}^{\ag} T_{\ag} U^{\bg} T_{\bg} \right)
      = K F_{ab}^{\ag} U^{\bg} \delta_{\ag \bg}.
      \label{eq:f}
\end{align}
For the Abelian group $U(1)$ there is only one generator and no auxiliary field
is needed.

Let us now apply the substitution~(\ref{eq:f}) to the above energy-momentum 
tensor~(\ref{eq:bi5})
\begin{multline}
      T_{ab} = \tr \left( F_{ac}^{\ag} T_{\ag} U^{\bg} T_{\bg} \right)
               \tr \left( F^{c}{}_{b}{}^{\ag} T_{\ag} U^{\bg} T_{\bg} \right)\\
               + \frac{1}{4}\eta_{ab}
               \tr \left( F_{cd}^{\ag} T_{\ag} U^{\bg} T_{\bg} \right)
               \tr \left( F^{cd\ag} T_{\ag} U^{\bg} T_{\bg} \right).
      \label{eq:bi6}
\end{multline}
In the following we will analyze the products of the two trace terms. For further
simplification we choose $\g=SU(2)$, but the principal result applies to any gauge
group $\g=SU(n)$; further note that the Lie algebra of $\g$ has $\dim(\g)=n^2-1$.
For $SU(2)$ there are three generators $T_{\ag}$, so explicitely we have
\begin{multline}
      \tr \left( F_{ac}^{\ag} T_{\ag} U^{\bg} T_{\bg} \right)
      \tr \left( F^{c}{}_{b}{}^{\ag} T_{\ag} U^{\bg} T_{\bg} \right) \\ =
      K^2 F_{ac}^{\ag} F^{c}{}_{b}{}^{\cg} U^{\bg} U^{\dg} 
      \delta_{\ag\bg} \delta_{\cg\dg}.
      \label{eq:bi7}
\end{multline}
If the introduced auxiliary field $U^{\ag}$ satisfies the following algebra
\begin{alignat}{3}
      U^{\ag} U^{\bg} = \frac{1}{K} \delta^{\ag \bg} ,
      \label{eq:bi8}
\end{alignat}
then Eq.~(\ref{eq:bi7}) becomes
\begin{multline}
      K^2 F_{ac}^{\ag} F^{c}{}_{b}{}^{\cg} \frac{1}{K} \delta^{\bg \dg} 
      \delta_{\ag\bg} \delta_{\cg\dg} \\ =
      K F_{ac}^{\ag} F^{c}{}_{b}{}^{\cg} \delta_{\ag \cg} = 
      \tr\left( F_{ac}^{\ag} T_{\ag} F^{c}{}_{b}{}^{\bg} T_{\bg} \right). 
\end{multline}
Hence, for the $SU(2)$ case the energy-momentum tensor~(\ref{eq:bi6}) takes
the standard form 
\begin{align}
      T_{ab} = \tr \left( F_{ac}^{\ag} T_{\ag} F^{c}{}_{b}{}^{\bg} T_{\bg}
      + \frac{1}{4}\eta_{ab} F_{cd}^{\ag} T_{\ag} F^{cd\bg} T_{\bg} \right).
      \label{eq:bi9}
\end{align}

The algebra~(\ref{eq:bi8}) of the auxiliary field $U^{\ag}$ has a simple geometrical
interpretation. Let $u^{\ag}$ be orthonormal basis of the vector space $\mathbb{R}^3$, then 
the three vectors $U^{\ag}=u^{\ag}/\sqrt{K}$ automatically satisfy the relations~(\ref{eq:bi8}). 
This can easily be generalized to $SU(n)$ by letting $u^{\ag}$ be the orthonormal 
vectors spanning the vector space $\mathbb{R}^{\dim(\g)}$ of the same dimension 
as the Lie algebra of $\g$. Choosing the auxiliary field $U^{\ag}=u^{\ag}/\sqrt{K}$ allows 
us to construct all $SU(n)$ Einstein-Yang-Mills field equations consistently.

\section{Remarks and Conclusions}

Firstly, it should be noted that one could have tried to extend the Kaluza-Klein 
scheme in a different manner, namely to allow $\dim(\g)$ extra 
dimensions rather than one extra dimension. However, such an approach does not 
yield the correct field equations by means of the contracted Bianchi identities, 
both in the torsion-free and the torsion case. Hence our auxiliary field approach
is, as far as we know, the only one giving correct results without the need to
introduce more than one additional spatial dimension.

The theory presented here is related to attempts to unify the four fundamental 
forces in a geometrical manner. However, for such an approach to work, 
requires the use of a gauge group that contains $U(1)$, $SU(2)$ and $SU(3)$
as subgroups, such as the Georgi-Glashow model that is based on $SU(5)$.
This model predicts a too fast proton decay rate and is therefore ruled out 
by experimental data. It, however, inspired other kinds of grand unification
theories based on a variety of gauge groups, like $SO(10)$ (double covering
of ${\rm Spin}(10)$) to mention the most prominent example, or the string
inspired and more complicated $E_6$ that also contains $SO(10)$. From that
point of view, our model can be seen as a geometrical argument in favor
of Grand Unified Theories.

Before summarizing our result, we would like to emphasize that we followed
two guiding principles: simplicity and agreement with experiment. The first
has been taken into account when we neglected the $z$-component of the 
connection 1-form, while agreement with experiment has led to the choice of
vanishing 4-dimensional torsion.

We showed that it is possible to extend the original Kaluza-Klein 
scheme to Yang-Mills theories. By considering a spacetime where the fifth 
dimension was allowed to carry torsion, we recovered the 4d
Einstein-Maxwell equations from the contracted Bianchi identities. 
We introduced a set of auxiliary fields which enabled us to obtain the 
Einstein-Yang-Mills field equations, also by considering the contracted Bianchi 
identities. The dynamics of the introduced auxiliary fields still need to be studied 
in detail. From a geometrical point of view this is a particularly interesting 
aspect of the approach since it does not require the prescription of an underlying 
action principle. It is geometry that dictates the form of the field equations,
contrary (but not in contradiction) to standard quantum field theories~\cite{Weinberg:1995mt}.
The key idea of this work, namely to introduce auxiliary fields rather than more 
and more additional dimensions to fit in the gauge fields, could be of special 
interest for a variety of today's models of theoretical physics.

\acknowledgments
We would like to thank Roy Maartens and Dmitri Vassiliev for valuable 
discussions. The work of CGB was supported by research grant BO 2530/1-1
of the German Research Foundation (DFG).

\end{document}